\documentclass[prb,showpaces,preprintnumbers,amsmath,amssymb,twocolumn,superscriptaddress,longbibliography]{revtex4-1}
\usepackage{graphicx}
\usepackage{dcolumn}
\usepackage{bm}
\usepackage{amssymb}
\usepackage{color}
\usepackage{amsmath}
\usepackage{epstopdf}
\usepackage{booktabs}
\usepackage{threeparttable}
\usepackage[pdfstartview=FitH, CJKbookmarks=true, bookmarksnumbered=true, bookmarksopen=true, colorlinks, pdfborder=001, linkcolor=blue, anchorcolor=blue, citecolor=blue]{hyperref}
\newcommand{\be}{\begin{equation}}
\newcommand{\ee}{\end{equation}}
\newcommand{\bea}{\begin{eqnarray}}
\newcommand{\eea}{\end{eqnarray}}
\newcommand{\nn}{\nonumber}

\begin{document}
\title{Evidence for triplet superconductivity near an antiferromagnetic instability in CrAs}
\author{C. Y. Guo}
\affiliation{Center for Correlated Matter and Department of Physics, Zhejiang University, Hangzhou 310058, China}
\author{M. Smidman}
\email{msmidman@zju.edu.cn}
\affiliation{Center for Correlated Matter and Department of Physics, Zhejiang University, Hangzhou 310058, China}
\author{B. Shen}
\affiliation{Center for Correlated Matter and Department of Physics, Zhejiang University, Hangzhou 310058, China}
\author{W. Wu}
\affiliation{Beijing National Laboratory for Condensed Matter Physics and Institute of Physics, Chinese Academy of Sciences, Beijing 100190, China}
\author{F. K. Lin}
\affiliation{Beijing National Laboratory for Condensed Matter Physics and Institute of Physics, Chinese Academy of Sciences, Beijing 100190, China}
\author{X. L Han}
\affiliation{Beijing National Laboratory for Condensed Matter Physics and Institute of Physics, Chinese Academy of Sciences, Beijing 100190, China}
\affiliation{University of Chinese Academy of Sciences, Beijing 100049, China}
\author{Y. Chen} 
\affiliation{Center for Correlated Matter and Department of Physics, Zhejiang University, Hangzhou 310058, China}
\author{F. Wu} 
\affiliation{Center for Correlated Matter and Department of Physics, Zhejiang University, Hangzhou 310058, China}
\author{Y. F. Wang} 
\affiliation{Center for Correlated Matter and Department of Physics, Zhejiang University, Hangzhou 310058, China}
\author{W. B. Jiang} 
\affiliation{Center for Correlated Matter and Department of Physics, Zhejiang University, Hangzhou 310058, China}
\author{X. Lu}
\affiliation{Center for Correlated Matter and Department of Physics, Zhejiang University, Hangzhou 310058, China}
\affiliation{Collaborative Innovation Center of Advanced Microstructures, Nanjing University, Nanjing 210093, China}
\author{J. P. Hu}
\affiliation{Beijing National Laboratory for Condensed Matter Physics and Institute of Physics, Chinese Academy of Sciences, Beijing 100190, China}
\affiliation{Collaborative Innovation Center of Quantum Matter, Beijing 100871, China}
\author{J. L. Luo}
\affiliation{Beijing National Laboratory for Condensed Matter Physics and Institute of Physics, Chinese Academy of Sciences, Beijing 100190, China}
\affiliation{Collaborative Innovation Center of Quantum Matter, Beijing, China}
\author{H. Q. Yuan}
\email{hqyuan@zju.edu.cn}
\affiliation{Center for Correlated Matter and Department of Physics, Zhejiang University, Hangzhou 310058, China}
\affiliation{Collaborative Innovation Center of Advanced Microstructures, Nanjing University, Nanjing 210093, China}
\date{\today}

\begin{abstract}
Superconductivity was recently observed in CrAs as the helimagnetic order is suppressed by applying pressure, suggesting possible unconventional superconductivity.  To reveal the nature of the superconducting order parameter  of CrAs, here we report the angular dependence of the upper critical field under pressure. Upon rotating the field by 360 degrees in the $bc$-plane, six maxima are observed in the upper critical field, where the oscillations have both six-fold and two-fold symmetric components.  Our analysis suggests the presence of an unconventional odd-parity spin triplet state.
\end{abstract}

\pacs{}
\maketitle

\section{Introduction}
Unconventional superconductivity, where the Cooper pairs are bound by a mechanism other than the conventional electron-phonon pairing mechanism described by BCS theory, has often been found to occur in close proximity to magnetic order in systems such as the cuprates, iron pnictides and heavy fermion superconductors.  For instance, in many heavy fermion materials, magnetism is suppressed to zero temperature at a quantum critical point (QCP) upon tuning with non-thermal parameters, which is often surrounded by a superconducting dome \cite{NatureMagSC,PfleidererRMP}. Non-Fermi liquid behavior is observed in the quantum critical region, indicating the presence of strong  critical spin fluctuations, which may mediate the unconventional superconductivity in these materials.

\begin{figure}[t]
\begin{center}
 \includegraphics[width=0.7\columnwidth]{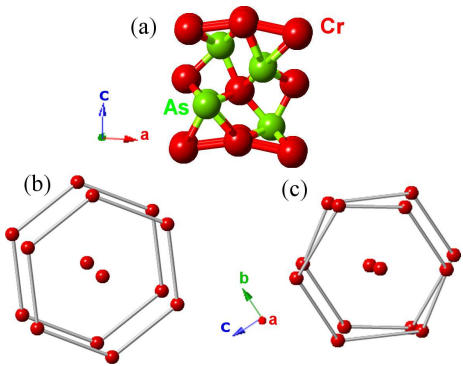}
\end{center}
\caption{ (a) Unit cell of CrAs at room temperature with an orthorhombic structure, where Cr atoms are shown in red and As in green. The arrangement of the Cr atoms are shown for (b) the  hexagonal NiAs-type structure and (c) the orthorhombic structure of CrAs which is a distortion of the aforementioned hexagonal structure. }
\label{Fig4}
\end{figure}

 Recently, some other $d$-electron compounds have also been reported to show similar phase diagrams to the Ce based heavy fermion superconductors, namely CrAs and MnP \cite{CrAsSC,CrAsSC2,MnPSC}.  CrAs and MnP  crystallize in an orthorhombic structure at room temperature, as displayed in Fig.~\ref{Fig4}(a). The room temperature crystal structure corresponds to a  distortion of the hexagonal NiAs-type structure with space group $P6_3/mmc$ and point group $D_{6h}$ to the orthorhombic phase with  space group $Pnma$ and point group $D_{2h}$ \cite{CrAsBook}. The configuration of the Cr atoms in the undistorted hexagonal structure is displayed in  Fig.~\ref{Fig4}(b), which is suggested to exist well above room temperature in CrAs \cite{watanabe1969magnetic}, where two layers of hexagonally arranged Cr atoms are shown, with one Cr atom in the centre.  The arrangement of the Cr atoms after the orthorhombic distortion  is shown in Fig.~\ref{Fig4}(c),  where the hexagons are distorted along one direction in the layer.   At ambient pressure CrAs shows a first-order magnetic transition at $T_N$~=~265~K to a helimagnetic state, which occurs concurrently with a structural transition where there is an elongation of the $b$-axis \cite{watanabe1969magnetic,seite1971magnetic,CrAsNeutP1,CrAsNeutP2}.  The ordering temperature $T_N$ is  suppressed upon applying pressure and vanishes above around 0.7~GPa \cite{CrAsSC,CrAsSC2}, which likely coincides with the disappearance of the structural transition \cite{CrAsNeutP2}. Meanwhile superconductivity under pressure is also observed in the highest quality samples, where at lower pressures there is phase separation between superconductivity and magnetism, while at higher pressures $T_N$ disappears and $T_c$  reaches a maximum of around 2~K \cite{CrAsSC,CrAsSC2,CrAsMuon}. Evidence for spin fluctuations in the normal state was also found from the non-Fermi liquid behavior of the resistivity with $\rho(T)\sim T^{1.5}$, as well as nuclear quadrupole resonance (NQR) measurements \cite{CrAsSCNQR,CrAsSpinFluc}. MnP also shows similar superconducting properties, where a superconducting dome is again observed after the suppression of magnetic order under pressure, which occurs at a higher pressure of around 8.0~GPa \cite{MnPSC}.

Another related family of Cr-based superconductors   $A_2$Cr$_3$As$_3$ ($A$~=~K, Rb or Cs) \cite{K2Cr3As3Rep,Rb2Cr3As3Rep,Cs2Cr3As3Rep} were  recently reported with a hexagonal crystal structure (space group $P\bar{6}m2$). The structure  shows some similarities to the undistorted hexagonal structure of CrAs, although rather than having a planar hexagonal arrangement, the Cr atoms form face sharing octahedra. Despite the lack of long range magnetic order, these materials also display a number of novel superconducting properties including very large and anisotropic upper critical fields  \cite{K2Cr3As3Crys,K2Cr3As3Hc2Big,zuo2015unconventional}, nodal superconducting order parameters \cite{K2Cr3As3Pen,K2Cr3As3MuSR,Rb2Cr3As3NMR} and low dimensional spin fluctuations in the normal state \cite{K2Cr3As3NMR,taddei2017coupling}. It is therefore of particular interest  to probe whether the superconductivity in CrAs and $A_2$Cr$_3$As$_3$ shares a common nature.

The characterization of pressure-induced superconductivity is experimentally challenging. Until now, very little is known about the superconducting gap symmetry and therefore the pairing mechanism of CrAs, even though its superconductivity appears on the border of antiferromagnetism. NQR measurements were  performed under pressure at temperatures down to about 1~K ($\approx T_c/2$), which show that the temperature dependence of the spin-lattice relaxation rate $1/T_1$ lacks a coherence peak and follows a $\sim T^3$ dependence below $T_c$ \cite{CrAsSCNQR}. Although these results indicate unconventional superconductivity in CrAs with possible line nodes in the energy gap, its pairing state is still unclear and further characterization of the superconducting order parameter is badly needed. The field-angle dependence of quantities such as the upper critical field ($B_{c2}$) and heat capacity are powerful tools which can provide important information about the symmetry and structure of the superconducting gap \cite{CCSHc2,CeCoIn5Hc2,CeCoIn5specH}. Here we report measurements of the  angular dependence of $B_{c2}$ of CrAs under pressure using a triple-axis vector magnet, in order to characterize the superconducting state of CrAs. We find a clear anisotropy of $B_{c2}$ in the $bc$~plane, where both two-fold and six-fold oscillatory components are observed, which are consistent with an odd-parity triplet superconducting state.

\section{Experimental Details}

 Single crystals of CrAs were synthesized using Sn-flux, as described previously \cite{CrAsGrowth}. The elements were combined in an atomic ratio  Cr:As:Sn of 3:3:40, placed in an alumina crucible and sealed in an evacuated quartz ampoule. The ampoule was held at 650$^\circ$C for 8 hours, then 1000$^\circ$C for 15 hours, before being slowly cooled down to 600$^\circ$C and centrifuged. The needle-like single crystals, with the $a$-axis along the needle direction are high quality, with a residual resistivity ratio at ambient pressure of $\rho(300~{\rm K})/\rho(2~{\rm K})\approx 330$.  Resistivity measurements under pressure were carried out using a piston-cylinder-type pressure cell, with Daphene 7373 used as a pressure transmitting medium to ensure  hydrostaticity. The sample was mounted onto a rectangular plastic plate so that the needle axis of the sample was aligned with the long direction of the plate. The plate was then attached vertically to the plug of the pressure cell. Although a small misalignment cannot be ruled out, this should not significantly affect our conclusions. The pressure was determined by measuring the superconducting transition of  Pb. The angular dependence of the magnetoresistance was measured using a 370 resistive bridge with a triple-axis vector superconducting magnet that accommodates a $^3$He cryostat. All the measurements were performed at one pressure of 1.3~GPa, with the current along the $a$-axis.

\begin{figure}[t]
\begin{center}
  \includegraphics[width=0.99\columnwidth]{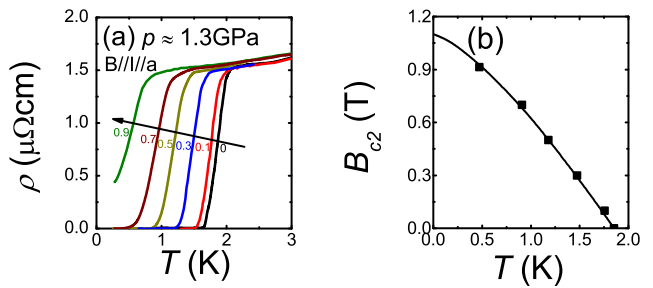}
\end{center}
	\caption{ (a) Temperature dependence of the electrical resistivity $\rho(T)$ at 1.3~GPa under various magnetic fields, with the field and current both applied parallel to the $a$-axis. (b) Temperature dependence of the upper critical field $B_{c2}(T)$,  determined from where there is a 50\% drop of the resistivity relative to the normal state. The solid line shows a fit to an empirical formula described in the text.}
   \label{Fig1}
\end{figure}

\section{Results}
\subsection{Temperature  and field-angle dependence of the resistivity}

\begin{figure*}[t]
\begin{center}
  \includegraphics[width=1.6\columnwidth]{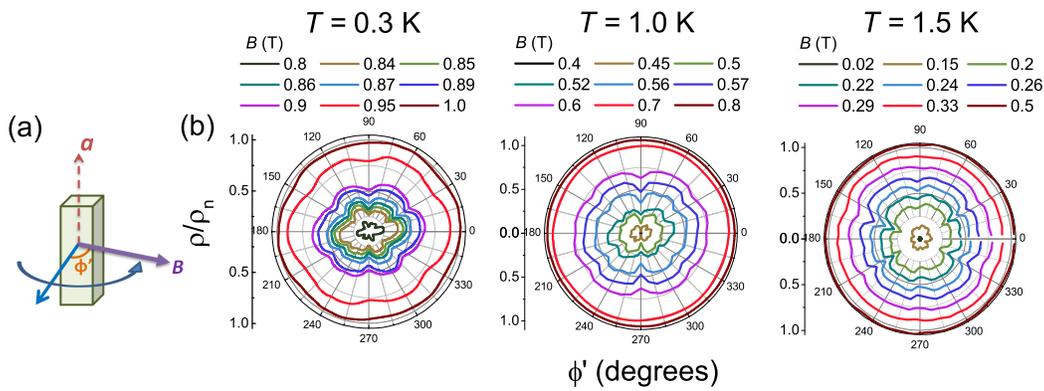}
\end{center}
	\caption{ (a) Illustration of the configuration for field-angle dependent resistivity measurements. (b) Field-angle dependence of the electrical resistivity $\rho(\phi')$ for various magnetic fields applied within the $bc$-plane at three temperatures, where the $\phi'$ is an azimuthal angle.}
   \label{Fig1b}
\end{figure*}

The temperature dependence of the resistivity ($\rho(T)$) of needle-shaped single crystals were measured with a current applied along the needle-direction, which corresponds to the $a$-axis. The measurements were performed at 1.3~GPa, well above the critical pressure for the suppression of magnetism and the structural transition, close to the pressure where the $T_c$ is maximum and the normal state is paramagnetic. As shown in Fig.~\ref{Fig1}(a),  a sharp superconducting transition is observed at about 1.85 K, slightly  higher than reported previously \cite{CrAsSC}. The residual resistivity is about 1.5 $\mu\Omega$cm, which is in good agreement with the former study and along with the sharp transition, indicates a high sample quality. Upon applying a field, the superconducting transition in $\rho(T)$ is suppressed to lower temperatures.  The values of $B_{c2}(T)$  are determined from where $\rho(T)$ drops to 50\% of the normal state value, and are displayed in  Fig.~\ref{Fig1}(b). The data are fitted using the empirical equation $B_{c2}$(T) = $B_{c2}$(0)$\{$1-[$T$/$T_c$]$^n$$\}$ , with $n$ = 1.36. The extrapolated zero temperature value $B_{c2}(0)$~=~1.1~T, is similar to the orbital limiting field ($B^{orb}_{c2}(0)$) of about 1.03~T calculated using $B^{orb}_{c2}~=~0.72T_c({\rm d}B_{c2}/{\rm d}T)_{T=T_c}$. This is  compared to a Pauli limiting field of $B_P=1.86T_c=3.44$~T and corresponds to a Maki parameter $\alpha_M=\sqrt{2}B^{orb}_{c2}(0)/B_P$ of about $\alpha_M=0.42$, which indicates that orbital pair breaking is the dominant pair breaking mechanism in an applied magnetic field, and Pauli limiting is not expected to be significant. We also measured the resistivity upon  rotating the applied magnetic field, so as to determine the angular dependence of $B_{c2}$ at different temperatures. In Fig.~\ref{Fig1b}, the resistivity is shown at three temperatures for various fields applied within the $bc$ plane, while the current was applied along the $a$~axis, always perpendicular to the field direction. It can be seen that there are six dips of the resistivity upon rotating the field through 360$^\circ$, which correspond to six maxima of $B_{c2}$.\\

\subsection{Angle-dependence of the upper critical field}

\begin{figure}[t]
\begin{center}
  \includegraphics[width=0.99\columnwidth]{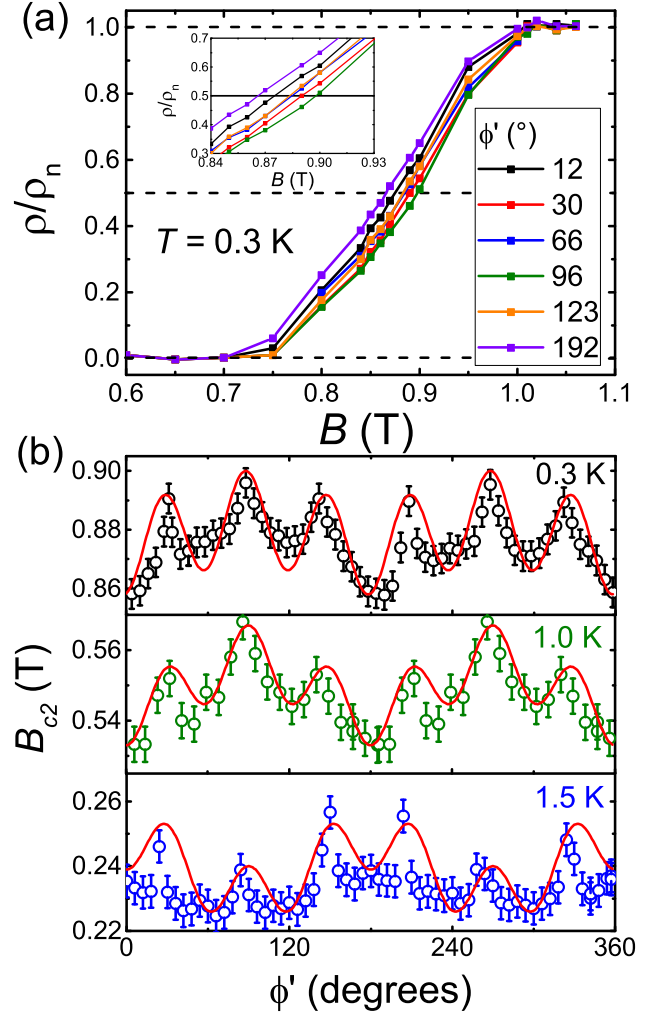}
\end{center}
	\caption{(a) Determination of the upper critical field  $B_{c2}$ for different field angles $\phi'$ within the $bc$-plane  at 0.3~K, from the magnetic field dependence of the resistivity normalized by the normal state value. The values of $B_{c2}$ were taken from where the resistivity reaches half the normal state value, as shown by the dashed line. The inset shows an enlargement of the data, in the vicinity of the upper critical fields.  (b) Angular dependence of $B_{c2}$  in the $bc$-plane at 0.3~K, 1~K and 1.5~K. The  lines show the fits for the angular dependence of an odd-parity spin triplet  state with both  two-fold and six-fold symmetric components. The error bars are determined from the size of the field increments for the angle-dependent resistivity measurements.}
   \label{Fig2}
\end{figure}

\begin{figure}[t]
\begin{center}
 \includegraphics[width=0.7\columnwidth]{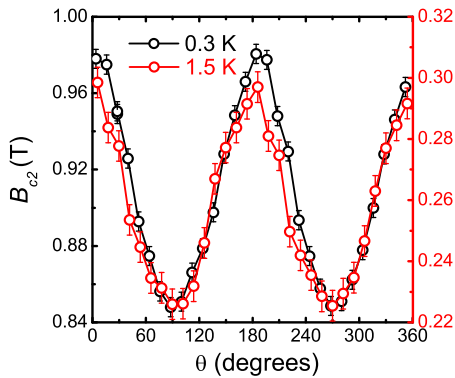}
\end{center}
\caption{ Angular dependence of the upper critical field $B_{c2}$($\theta$) as a function of the polar angle $\theta$, where $\theta=0$ corresponds to the $a$-axis. }
\label{Fig3}
\end{figure}

The angular dependence of the upper critical field in the $bc$-plane [$B_{c2}$($\phi'$)] is shown in Fig.~\ref{Fig2}. Note that due to  the thin needle-like shape of the single crystals, the position within the $bc$-plane from which the azimuthal angle $\phi'$  is measured could not be determined, and therefore $\phi'=0$ was chosen to correspond to where $B_{c2}$ is at a minimum. The $B_{c2}$ values were obtained by measuring $\rho(\phi')$ in various fields and taking the field where $\rho$ is half  the normal state value, as displayed in Fig.~\ref{Fig2}~(a). As shown in Fig.~\ref{Fig2}~(b), at all temperatures there are six maxima in $B_{c2}$($\phi'$) which are approximately  separated by 60$^\circ$. Overall $B_{c2}$($\phi'$) also shows the two-fold symmetry of the orthorhombic crystal structure of CrAs and therefore these results suggest the presence of both six-fold and two-fold symmetric components to the oscillations. There are also differences in the intensities of the peaks. At 0.3~K, the largest peaks are at about 90$^\circ$ and 270$^\circ$ and upon increasing the temperature to 1.0 K, the difference between the intensities of these two peaks and the others is enlarged. However at 1.5~K, the peaks at 150$^\circ$ and 210$^\circ$ now have the greatest intensity, in contrast to the measurements at 0.3~K and 1.0~K. This temperature is very close to $T_c$ where the angular variation is much weaker, whereas at lower temperatures the anisotropy is much more pronounced. The presence of these six-fold and two-fold components within the $bc$-plane indicate that there is a non-$s$-wave  pairing state. The out-of-plane angular dependence of the upper critical field as a function of the polar angle [$B_{c2}$($\theta$)] is shown in Fig.~\ref{Fig3}, where two large peaks can be observed rotating through 360$^\circ$. The peak positions  are at about 0$^\circ$ and 180$^\circ$, which correspond to fields along the $a$-axis.

The observation of an oscillatory $B_{c2}$($\phi'$) gives evidence for  unconventional superconductivity in CrAs. In the following, we discuss the potential superconducting states compatible with this observation. We  first derive the relation between the pairing symmetry  and the angular dependence of the upper critical field $H_{c2}$ based on  Gorkov's theory \cite{GorkovGL,WHH2,Mineev1998}. Starting from a Hamiltonian with electron-electron interactions,

\begin{multline}
H=\sum_{\sigma}\int d^3{\bf r}\bar{\psi}_{\sigma}({\bf r})(-\frac{\nabla^2}{2m}-\mu)\psi({\bf r}) \\+\int d^3{\bf r}\int d^{3} {\bf r}^{\prime} \bar{\psi}_{\uparrow}({\bf r})\bar{\psi}_{\downarrow}({\bf r}^{\prime})V(r-r^{\prime})\psi_{\downarrow}({\bf r}^{\prime})\psi_{\uparrow}({\bf r})
\end{multline}
where $m$ and $\sigma$ label the mass and spins of the electrons respectively, the equations of motion can be written using the normal and anomalous Green's functions as
\begin{multline} \label{GF}
\Big(i\omega_n-\frac{1}{2m}(-i\nabla+e {\bf A})^2\Big)G_{\sigma,\sigma^{\prime}}({\bf r},{\bf r}^{\prime},\omega_n)\\+\sum_{\rho}\int d^3 \xi \Delta_{\sigma\rho} ({\bf r},\xi)\bar{F}_{\rho\sigma^{\prime}}(\xi,{\bf r}^{\prime},\omega_{n})=\delta({\bf r}-{\bf r}^{\prime})\delta_{\sigma\sigma^{\prime}} 
\end{multline}
\begin{multline} 
\Big(-i\omega_n-\frac{1}{2m}(-i\nabla-e {\bf A})^2\Big)\bar{F}_{\sigma,\sigma^{\prime}}({\bf r},{\bf r}^{\prime},\omega_n)\\+\sum_{\rho}\int d^3 \xi \Delta_{\sigma\rho}^{\ast} ({\bf r},\xi)G_{\rho\sigma^{\prime}}(\xi,{\bf r}^{\prime},\omega_{n})=0
\end{multline}
with the two Green's functions defined as
\begin{equation}\label{G0}
\begin{split} 
G_{\sigma\sigma^{\prime}}({\bf r},{\bf r}^{\prime},\tau)=-i\Big\langle T_{\tau}\big(\psi_{\sigma}({\bf r},\tau)\bar{\psi}_{\sigma^{\prime}}({\bf r}^{\prime},0)\big)\Big\rangle,\\ \quad F({\bf r},{\bf r}^{\prime},\tau)=-i\Big\langle T_{\tau}\big(\psi_{\sigma}({\bf r},\tau)\psi_{\sigma^{\prime}}({\bf r}^{\prime},0)\big)\Big\rangle
\end{split}
\end{equation}
in which  $\psi(\tau)\equiv e^{H\tau}\psi(0)e^{- H\tau}$ is defined in the imaginary time domain. The order parameter is given by $\Delta_{\sigma\sigma^{\prime}}({\bf r},{\bf r}^{\prime})=V({\bf r}-{\bf r}^{\prime})F_{\sigma\sigma^{\prime}}({\bf r},{\bf r}^{\prime})$. The free Green's function $G^{0}$ satisfies

\begin{multline}
\Big(i\omega_n-\frac{1}{2m}(-i\nabla+e {\bf A})^2\Big)G_{\sigma,\sigma^{\prime}}^0({\bf r},{\bf r}^{\prime},\omega_n)\\=\delta({\bf r}-{\bf r}^{\prime})\delta_{\sigma\sigma^{\prime}}.
\end{multline}

\noindent From Eqs.~\ref{GF} and \ref{G0} we obtain the gap equation  as
\begin{multline}
\Delta_{\sigma\sigma^{\prime}}({\bf r},{\bf r}^{\prime})= T V({\bf r}-{\bf r}^{\prime})\\ \times\sum_{n,\rho}\int d^3 \xi d^3 \xi^{\prime}G^{0}_{\sigma^{\prime}\sigma^{\prime}}({\bf r}^{\prime},\xi^{\prime},-\omega_n) \Delta_{\sigma\rho}(\xi,\xi^{\prime})G_{\rho\sigma^{\prime}}({\bf r},\xi,\omega_n),
\label{gap2}
\end{multline}
where $T$ is the temperature and $V$ is the interaction strength. In the presence of a magnetic field, we can further simplify the above equation by introducing 
 \begin{eqnarray}
 G({\bf r},{\bf r}^{\prime})=e^{-ie\int_{{\bf r}^{\prime}}^{\bf r}A({\bf l})\cdot d {\bf l}}\overline{G}^0({\bf r},{\bf r}^{\prime}),
 \end{eqnarray}
and the corresponding $\overline{G}^0$, which satisfies
\bea
\Big(i\omega_n-\frac{1}{2m}(-i\nabla)^2\Big)\overline{G}_{\sigma,\sigma^{\prime}}^0({\bf r},{\bf r}^{\prime},\omega_n)=\delta({\bf r}-{\bf r}^{\prime})\delta_{\sigma\sigma^{\prime}},
\eea
with a solution
\begin{multline}
\overline{G}^{0}_{\sigma\sigma}({\bf r},{\bf r}^{\prime},\omega_n)=-\frac{m}{2\pi|{\bf r}-{\bf r}^{\prime}|}\exp\Big\{|{\bf r}-{\bf r}^{\prime}|\big[ip_{F} sgn(\omega_n)\\ -\frac{|\omega_n|}{v_F}\big]\Big\},
\end{multline}
where $p_{F}$ is the Fermi momentum. Upon plugging the above solution into Eq.~\ref{gap2} and considering the gap function in the center of mass coordinates and relative coordinates, we finally obtain the gap function 
%\begin{widetext}
%\bea\label{gapEq}
\begin{multline}
\Delta_{\sigma\sigma^{\prime}}({\bf R},{\bf k})=\int\frac{d^3 {\bf k}^{\prime}}{(2\pi)^3} T V({\bf k}-{\bf k}^{\prime}) \sum_{n,\rho}\int d^3 \xi d^3 \xi^{\prime} e^{i{\bf k}^{\prime}\cdot(\xi^{\prime}-\xi)}\\ \times\overline{G}^{0}_{\sigma^{\prime}\sigma^{\prime}}(\xi,-\omega_n) e^{i\xi\cdot \Pi({\bf R})} \Delta_{\sigma\rho}({\bf R},{\bf k}^{\prime})\overline{G}_{\rho\sigma^{\prime}}({\bf R},\xi^{\prime},\omega_n),
\label{gap5}
\end{multline}
%\eea
%\end{widetext}
in which  $\Pi({\bf R})\equiv -i\nabla+2e{\bf A}(\bf R)$ and  $\Delta({\bf R},{\bf k})=\int d^{3}{\bf r} e^{-i{\bf k}\cdot {\bf r}}\Delta({\bf R},{\bf r})$.

First, we take the in-plane $p$-wave pairing superconducting state \cite{Abrikosov,Kurt1980,Klemm1985,Carbotte1990,Wu1998} as an example, so as to derive the periodicity of $H_{c2}$. The $p$-wave order parameter is   $\Delta({\bf R},{\bf k})=\Delta({\bf R})\cos(\phi)(|\uparrow\downarrow\rangle+|\downarrow\uparrow\rangle)/\sqrt{2}$, where $\phi$ is the azimuthal angle.  In the weak coupling limit, Eq.~\ref{gap5} simplifies to

%\begin{widetext}
%\bea
\begin{multline}
\Delta({\bf R})=gT \sum_n\int d^3 \vec{\xi} \cos^2(\phi)\overline{G}^{0}_{\downarrow\downarrow}(\xi,-\omega_n)\\ \times \overline{G}_{\uparrow\uparrow}^{0}(\xi,\omega_n)e^{i\xi\cdot \Pi({\bf R})} \Delta({\bf R}).
\label{gap6}
\end{multline}
%\eea
%\end{widetext}

where $g$ is the interaction strength. By considering a magnetic field applied along the $c$-axis, $H_{c2}$ is determined by $\kappa\eta-\chi^{\ast}\chi=0$, where
\begin{widetext}
\bea
\kappa&&=1-gT\sum_n\int d^3\vec{\xi} \cos^2(\phi)\overline{G}^{0}_{\downarrow\downarrow}(\xi,-\omega_n) \overline{G}_{\uparrow\uparrow}^{0}(\xi,\omega_n)e^{-\frac{eH_{c2} \xi^2}{2}}, \\
\chi&&=-gT\sum_n\int d^3 \vec{\xi} \cos^2(\phi) e^{2i\phi} \overline{G}^{0}_{\downarrow\downarrow}(\xi,-\omega_n) \overline{G}_{\uparrow\uparrow}^{0}(\xi,\omega_n)e^{-\frac{eH_c \xi^2}{2}}\frac{e H_{c2} \xi^2 \sin^2\theta}{\sqrt{2}},\\
\eta&&=1-gT\sum_n\int d^3 \vec{\xi} \cos^2(\phi)\overline{G}^{0}_{\downarrow\downarrow}(\xi,-\omega_n) \overline{G}_{\uparrow\uparrow}^{0}(\xi,\omega_n)e^{-\frac{eH_c \xi^2}{2}}L_2(e H_{c2}\xi^2\sin^2\theta).
\eea
\end{widetext}
Here $\theta$ is the polar angle  and $L_2(x)$ is the second Laguerre polynomial, which is equal to $(x^2-4x+2)/2$.

When  there is an in-plane magnetic field at an angle $\theta_0$ to  the $a$-axis, we can rotate the coordinates and rewrite the above equations by replacing ${\rm cos}^2(\phi)$ with the function
\bea
M(\theta_0,\vec{\xi^{\prime}}).=\frac{(-\sin\theta_0 \xi^{\prime}_{y^{\prime}}+\cos\theta_0 \xi^{\prime}_{z^{\prime}})^2}{(\xi^{\prime})^2-(\xi^{\prime}_{z^{\prime}})^2},
\eea
so that Eq.~\ref{gap6} becomes 
%\begin{widetext}
%\bea\label{pgap}
\begin{multline}
\Delta({\bf R})=gT\sum_n\int d^3 \vec{\xi} M(\theta_0,\vec{\xi})\overline{G}^{0}_{\downarrow\downarrow}(\xi,-\omega_n)\\ \times \overline{G}_{\uparrow\uparrow}^{0}(\xi,\omega_n)e^{i\xi\cdot \Pi({\bf R})} \Delta({\bf R}).
\label{gap7}
\end{multline}
%\eea
%\end{widetext}
It is very clear from these equations that $H_{c2}$ must be a function of $\cos(2\theta_0)$ and $\sin(2\theta_0)$,  with an angular periodicity of $\pi$.  Considering that the angular dependence is generally weak, we must have 
\begin{eqnarray}
H_{c2}(\theta_0)=H^0_{c2}+\delta H_{c2} \cos(2\theta_0+\alpha'),
\end{eqnarray}
where $\alpha'$ is a constant for a given model.

The above result can immediately be extended to other pairing states. In general, if the superconducting order parameter has  an in-plane gap function that is proportional  to $\cos(n\phi_0)$,  $H_{c2}$ must have an angular period equal to $\frac{\pi}{n}$, namely, 
\begin{eqnarray}
H_{c2}(\theta_0)=H^0_{c2}+\delta H_{c2} \cos(2n\theta_0+\alpha').
\label{GenHc2}
\end{eqnarray}

The crystal structure of CrAs is orthorhombic with space group $Pnma$ and point group $D_{2h}$.   As the orthorhombic distortion is rather small, we can characterize the superconducting properties by the   $P6_3/mmc$ group together with a perturbative deviation caused by the orthorhombic distortion.  Ignoring the distortion, the pairing symmetry should be classified by the $D_{6h}$ point group. The hexagonal pattern of the in-plane anisotropy is consistent with the odd-parity  spin triplet $f$-wave pairing symmetry, which belongs to the $B_{1u}$ or $B_{2u}$ irreducible representation of $D_{6h}$. Assuming $\Delta({\bf R},{\bf k})=\Delta({\bf R})\Delta(\phi)(|\uparrow\downarrow\rangle+|\downarrow\uparrow\rangle)/\sqrt{2}$, and taking  the $B_{1u}$ gap function as an example, the specific gap function is proportional to $x(x^2-3y^2)={\rm cos}(3\phi)$. Under such a pairing symmetry, the anisotropy of the in-plane $B_{c2}$ is given by

\begin{eqnarray}
B_{c2}(\phi')= \alpha+\beta {\rm cos}^2(3\phi'),
\label{Bc2Hex}
\end{eqnarray}

\noindent which leads to a  six-fold symmetry of the anisotropy following Eq.~\ref{GenHc2}, where $\phi'=0$ is chosen so that $\alpha'=0$. We note that the only other pairing states compatible with the in-plane anisotropy are the spin singlet $B_{1g}$ or $B_{2g}$ states. However these higher order pairings are expected to be energetically unfavourable, owing to the additional gapless region for $k_z=0$.

Moreover, we can consider an orthorhombic distortion to the hexagonal structure where the point group symmetry is lowered to $D_{2h}$, which corresponds to CrAs in the superconducting state. Taking the $B_{1u}$ gap function $x(x^2-3y^2)$ under $D_{6h}$, both terms in $x(x^2-3y^2)$, namely, $x^3$ and $xy^2$, belong to $B_{3u}$ under $D_{2h}$. Therefore for the orthorhombic lattice,  the lattice distortion can lead to the inclusion of a general symmetry broken term as 
\bea\label{p-wave}
\Delta(\phi)&&=x(x^2-3y^2)+\delta(x^3+3x y^2) \nn \\
&&=\Delta_0\cos(3\phi)+\delta \cos(\phi)+2\delta \cos(\phi)\sin^2(\phi).
\eea
It is reasonable to neglect the last term in Eq.~\ref{p-wave}, since $\delta$ is much smaller than $\Delta_0$.  Thus the pairing symmetry is approximately given by 
\bea
\Delta(\phi)=\Delta_0\cos(3\phi)+\delta \cos(\phi),
\label{fp}
\eea
\noindent which indicates that  the superconducting state has a mixture of $f$-wave and $p$-wave pairing symmetries.  Thus  the anisotropy of the in-plane $B_{c2}$ can be generally written as  

\begin{eqnarray}
B_{c2}(\phi')= \alpha+\beta {\rm cos}^2(3\phi')+\gamma {\rm cos}^2(\phi').
\label{FitEq1}
\end{eqnarray}

\noindent As displayed in Fig.~\ref{Fig2}(b) this expression can account for the observed anisotropy in the $bc$-plane, indicating that an odd-parity triplet superconducting state best explains the observed anisotropy. A similar modulation is obtained for the  $B_{2u}$ irreducible representation of $D_{6h}$, but since we cannot determine the crystallographic direction from which $\phi'$ is measured, we are unable to distinguish between the $B_{1u}$ and $B_{2u}$ states.  At the lowest temperature, the six-fold component has a larger amplitude, with $\beta/\gamma\approx3$, while at the higher temperatures they have comparable magnitudes with  $\beta/\gamma\approx1$.

 The out-of-plane anisotropy in Fig.~\ref{Fig3} can also be explained by assuming that  the $S_z=\pm1$ triplet pairing is slightly favored over $S_z=0$ triplet pairing.  In this case, the magnetic field parallel to  the $z$ direction does not cause the Zeeman splitting to break the Cooper pairs while an in-plane magnetic field has a strong Zeeman effect to break them. However, we also note that the maximum values correspond to fields parallel to the current, while the minima correspond to when the field and current are perpendicular. Since the Lorentz force exerted on vortices follows $\mathbf{H}\times\mathbf{I}$, where  $\mathbf{H}$ and  $\mathbf{I}$ are the applied fields and currents, this can lead to a lowering of the upper critical field measured using the resistivity and therefore may account for the out of plane anisotropy. On the other hand, the measurements in the $bc$-plane were all performed with   $\mathbf{H}\perp\mathbf{I}$ and therefore the effect of the Lorentz force cannot account for the in-plane anisotropy. We also note that although the hexagonal cross-section of the needle-like single crystals may also give rise to a six-fold oscillation of the upper critical field due to demagnetization effects, this would require the system to have a significant magnetization. Since at the upper critical field the magnetization approaches the normal state value, which likely corresponds to Pauli paramagnetism \cite{JAP1993}, and there is not a significant variation of the demagnetization factor within the $bc$-plane, demagnetization effects do not account for the observed in-plane modulation. On the other hand, due to the thin needle-like shape of the sample, the change of $B_{c2}$  as a function of $\theta$ (Fig.~\ref{Fig3}) could have a contribution from demagnetization effects.

\section{Discussion}

The presence of an odd-parity spin triplet state deduced from an oscillatory  $B_{c2}$ within the $bc$-plane is also consistent with the evidence for line nodes and lack of coherence peak below $T_c$ from NQR measurements \cite{CrAsSCNQR}. Furthermore, the angular dependence of $B_{c2}$ of CrAs is remarkably similar to that of K$_2$Cr$_3$As$_3$  \cite{zuo2015unconventional}, which shows a similar six-fold oscillation in the $ab$-plane. This suggests $f$-wave pairing, which is one of the candidates proposed theoretically \cite{K2Cr3As3Theor1,K2Cr3As3Theor2}. Therefore the common features in the upper critical field and structural similarities between these Cr-based superconductors, suggest a close relationship between the unconventional superconducting states of these materials, and that they may share a similar pairing mechanism. Moreover, there is evidence for spin fluctuations in the  normal states of both compounds. In the case of K$_2$Cr$_3$As$_3$  both nuclear magnetic resonance and neutron scattering measurements indicate that there are short range antiferromagnetic spin fluctuations   \cite{K2Cr3As3NMR,taddei2017coupling}. Meanwhile the presence of spin fluctuations in CrAs is inferred from the low temperature increase of $1/T_1T$ from NQR measurements, where the suppression of both  $1/T_1T$ and $T_c$ with increasing pressure suggests that an important role is played by  spin fluctuations in the formation of superconductivity \cite{CrAsSCNQR,CrAsSpinFluc}. Together with our results, this could indicate the occurence of triplet  superconductivity mediated by antiferromagnetic spin fluctuations. This may be similar to the putative spin triplet superconductor UPt$_3$, which also shows very weak antiferromagnetic order below $T_N=5$~K \cite{UPt3AFM1,UPt3NS,UPt3Rev}, whereas antiferromagnetic correlations set in at higher temperatures \cite{AEPPLI1988385,UPt3Rev}. Therefore how such antiferromagnetic fluctuations can lead to a spin triplet pairing state requires further experimental and theoretical exploration.

A further similarity between CrAs and  UPt$_3$ is that both the undistorted hexagonal arrangement and the orthorhombic crystal structure of CrAs correspond to nonsymmorphic crystal structures. In the case of UPt$_3$, it was shown that non-symmorphic symmetry allowed for the existence of  triplet superconductivity with line nodes \cite{Norman1995},  and such symmetries have been shown more generally to lead to the topological protection of gap nodes \cite{Kobayashi2016}.  Nonsymmorphic symmetries have also been found to lead to a variety of unusual topological states in condensed matter systems \cite{wang2016hourglass,ma2017experimental,chang2017mobius}, and it was recently predicted that in UPt$_3$, the glide symmetry leads to a topological M\"obius superconducting state with double Majorana cone surface states \cite{TopoUPt3}. Meanwhile, a recent magnetotransport study of the normal state of CrAs found a non-saturating linear magnetoresistance under pressure, which was suggested to originate from the topologically non-trivial band crossing protected by the nonsymmorphic symmetry \cite{CrAsQLMR}. Importantly, band structure calculations suggest that this novel band crossing is very close to Fermi level, and it is therefore of great interest to determine in future whether these topological features influence the superconductivity of CrAs. Furthermore, given that topological superconductivity has been proposed to occur in nodal spin-triplet superconductors, which can lead to unusual surface states \cite{topoSCrev}, the possibility of its realization in CrAs due to our proposed triplet state with line nodes needs to be explored.

\section{Conclusions}

To summarize, we have measured the angular dependence of the upper critical field in the pressure-induced superconducting state of CrAs, which shows a six-fold oscillation in the $bc$-plane. These findings are explained most naturally by an odd-parity spin triplet pairing state. Furthermore, it is found that CrAs and K$_2$Cr$_3$As$_2$ display a similar angular dependence  of the in-plane upper critical field, suggesting that these Cr-based superconductors may share a similar mechanism for the formation of superconductivity. Further experimental and theoretical investigations are necessary  to reveal the unconventional pairing mechanism and explore the possibility of topological superconductivity in Cr-based superconductors.

\begin{acknowledgments}
We would like to thank C. Cao,  Z. Zhu,  S. Kirchner,  F. Steglich, D. F. Agterberg and F. -C. Zhang  for valuable discussions. This work was supported by the National Key R\&D Program of China (No.~2017YFA0303100, No.~2016YFA0300202, No.~2015CB921300), the National Natural Science Foundation of China (No.~U1632275, No.~11474251, No.~1190020, No.~11534014, No.~11334012) and the Science Challenge Project of China (No.~TZ2016004).
\end{acknowledgments}

\end{document}